\documentclass[runningheads]{llncs}
\usepackage{graphicx}
\usepackage{amsmath,amssymb} 
\usepackage{color}
\usepackage{wrapfig}
\usepackage{pifont}
\usepackage{subcaption}

\begin{document}
\pagestyle{headings}
\mainmatter
\def\ECCVSubNumber{195}  

\title{PointFISH: Learning Point Cloud Representations for RNA Localization Patterns} 

\titlerunning{PointFISH}
\authorrunning{A. Imbert et al.}
\author{Arthur Imbert$^{1, 2, 3, \dagger}$, Florian Mueller$^{4, 5}$, Thomas Walter$^{1, 2, 3, \dagger}$}
\institute{
1. Centre for Computational Biology, Mines Paris, PSL University, Paris, France\\
2. Institut Curie, PSL University, Paris, France\\
3. INSERM, U900, Paris, France\\
4. Imaging and Modeling Unit, Institut Pasteur and UMR 3691 CNRS, Paris, France\\
5. C3BI, USR 3756 IP CNRS, Paris, France\\
$\dagger$ Corresponding authors: Thomas Walter (Thomas.Walter@minesparis.psl.eu), Arthur Imbert (Arthur.Imbert@minesparis.psl.eu)}

\maketitle

\begin{abstract}
Subcellular RNA localization is a critical mechanism for the spatial control of gene expression.
Its mechanism and precise functional role is not yet very well understood.
Single Molecule Fluorescence in Situ Hybridization (smFISH) images allow for the detection of individual RNA molecules with subcellular accuracy. 
In return, smFISH requires robust methods to quantify and classify RNA spatial distribution.
Here, we present PointFISH, a novel computational approach for the recognition of RNA localization patterns.
PointFISH is an attention-based network for computing continuous vector representations of RNA point clouds.
Trained on simulations only, it can directly process extracted coordinates from experimental smFISH images.
The resulting embedding allows scalable and flexible spatial transcriptomics analysis and matches performance of hand-crafted pipelines.
\end{abstract}

\section{Introduction}
\label{sec:introduction}

Localization of messenger RNAs (mRNAs) are of functional importance for gene expression and in particular its spatial control.
RNA localization can be related to RNA metabolism (to store untranslated mRNAs or degrade them) or protein metabolism (to localize translations).
RNA localization is not a limited phenomenon but a common mechanism throughout the transcriptome, which might also concern non-coding RNAs~\cite{lecuyer_global_2007,buxbaum_right_2015}.
Despite the importance of this process, it is still poorly understood, and adequate tools to study this process are still lacking.

\begin{figure}[]
	\centering
	\minipage{0.2\textwidth}
		\includegraphics[width=0.95\linewidth]{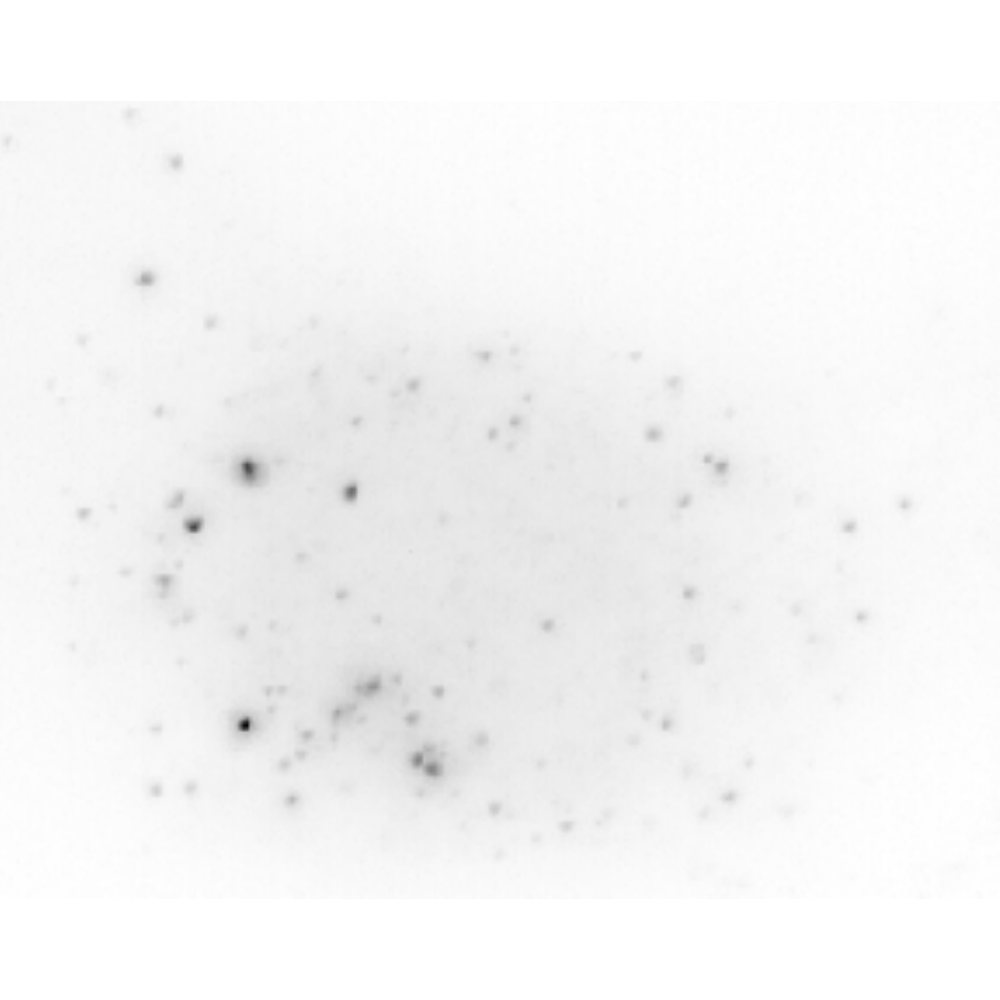}
		\vfill
		\includegraphics[width=0.95\linewidth]{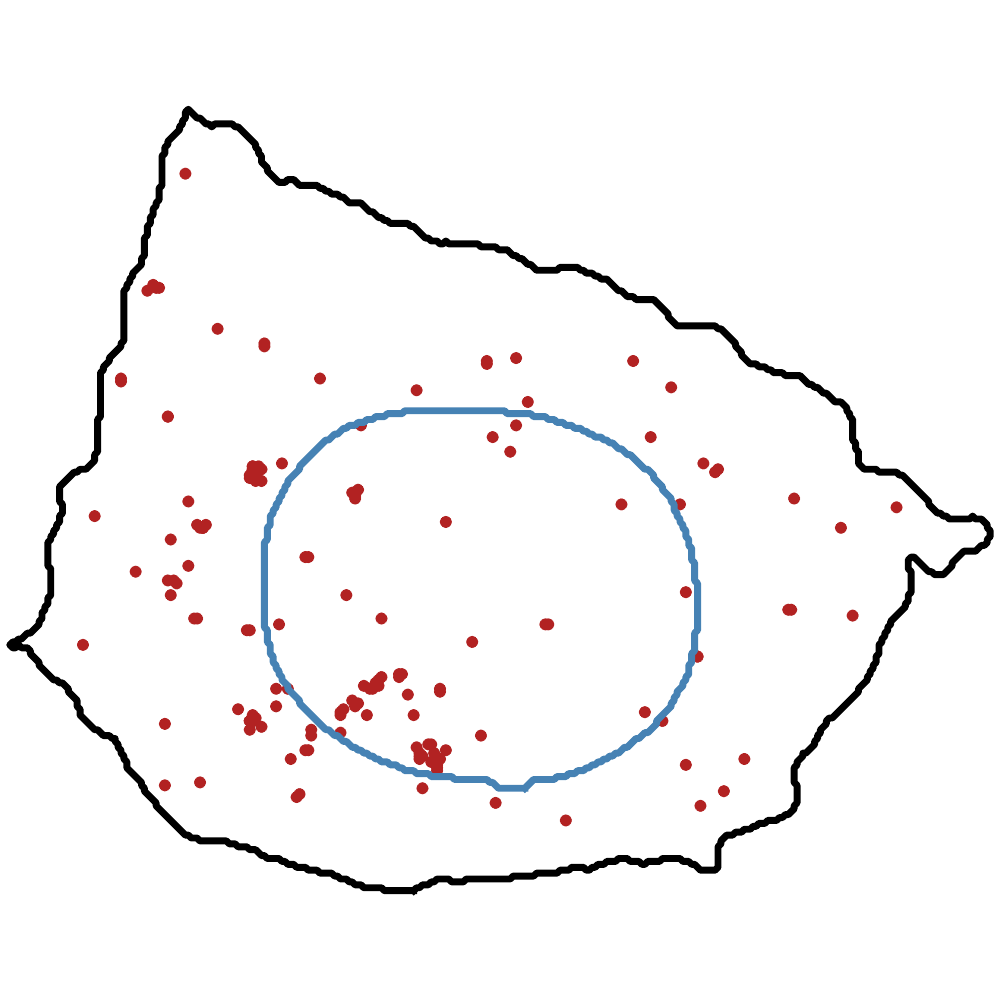}
		\subcaption{Foci}
	\endminipage\hfill
	\minipage{0.2\textwidth}
		\includegraphics[width=0.95\linewidth]{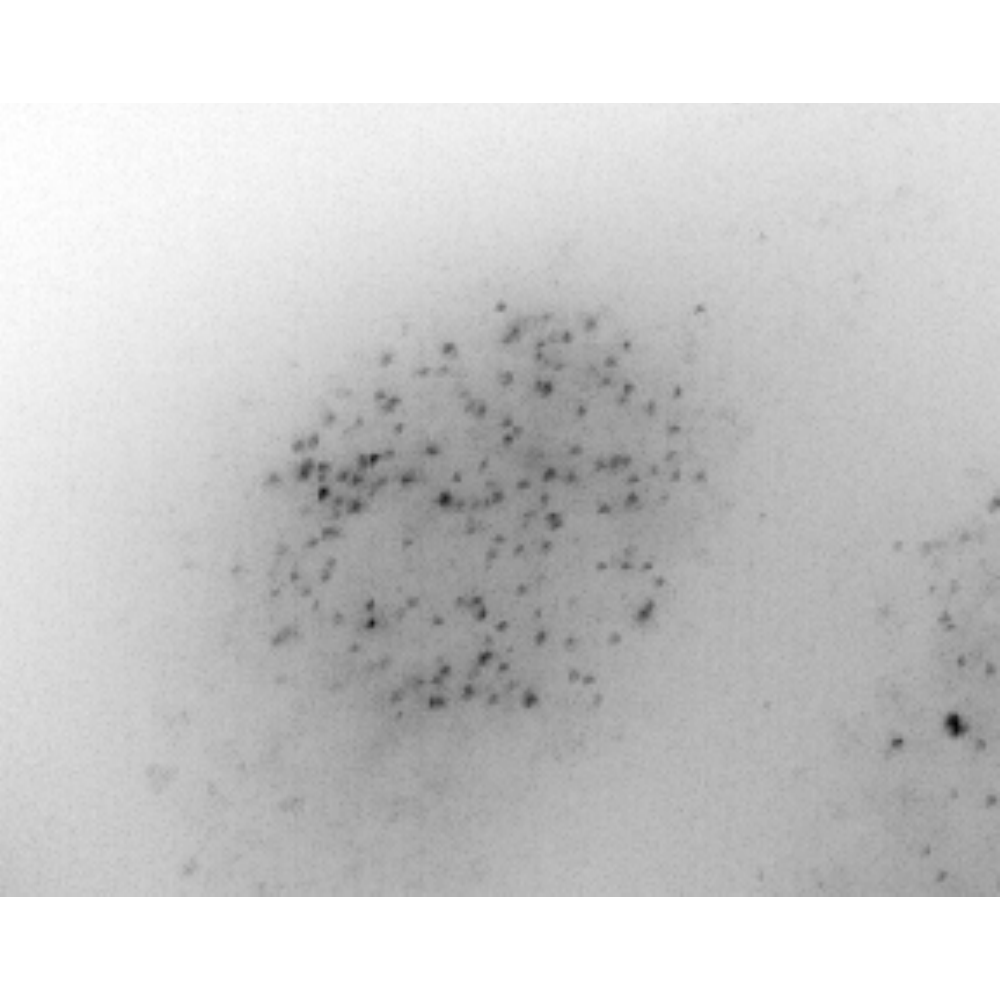}
		\vfill
		\includegraphics[width=0.95\linewidth]{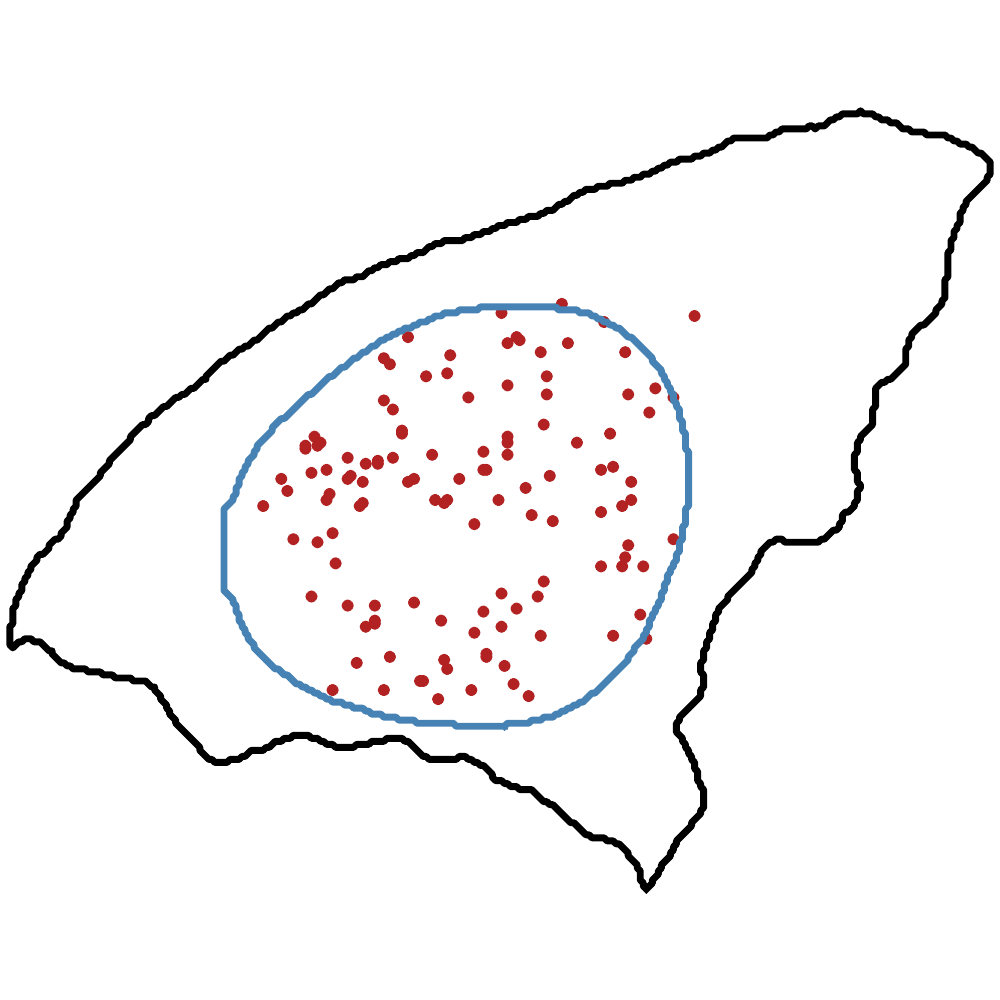}
		\subcaption{Intranuclear}
	\endminipage\hfill
	\minipage{0.2\textwidth}
		\includegraphics[width=0.95\linewidth]{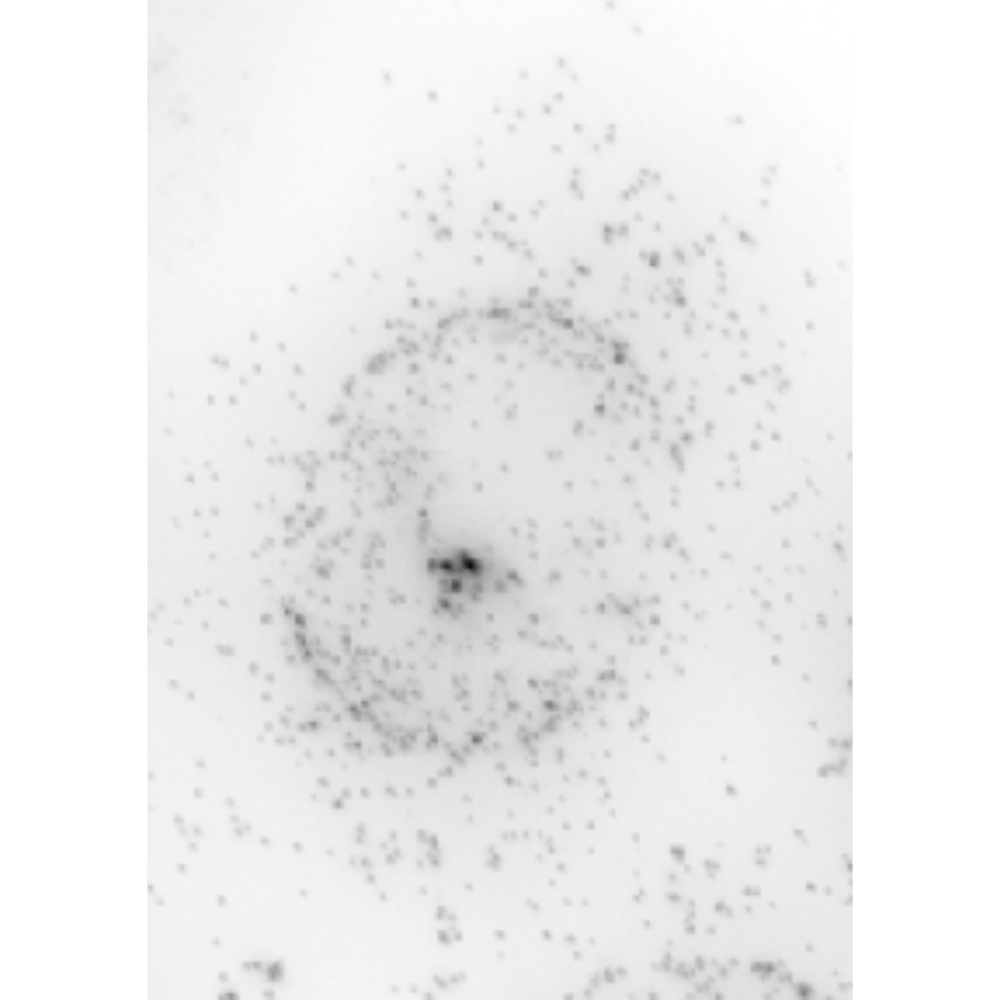}
		\vfill
		\includegraphics[width=0.95\linewidth]{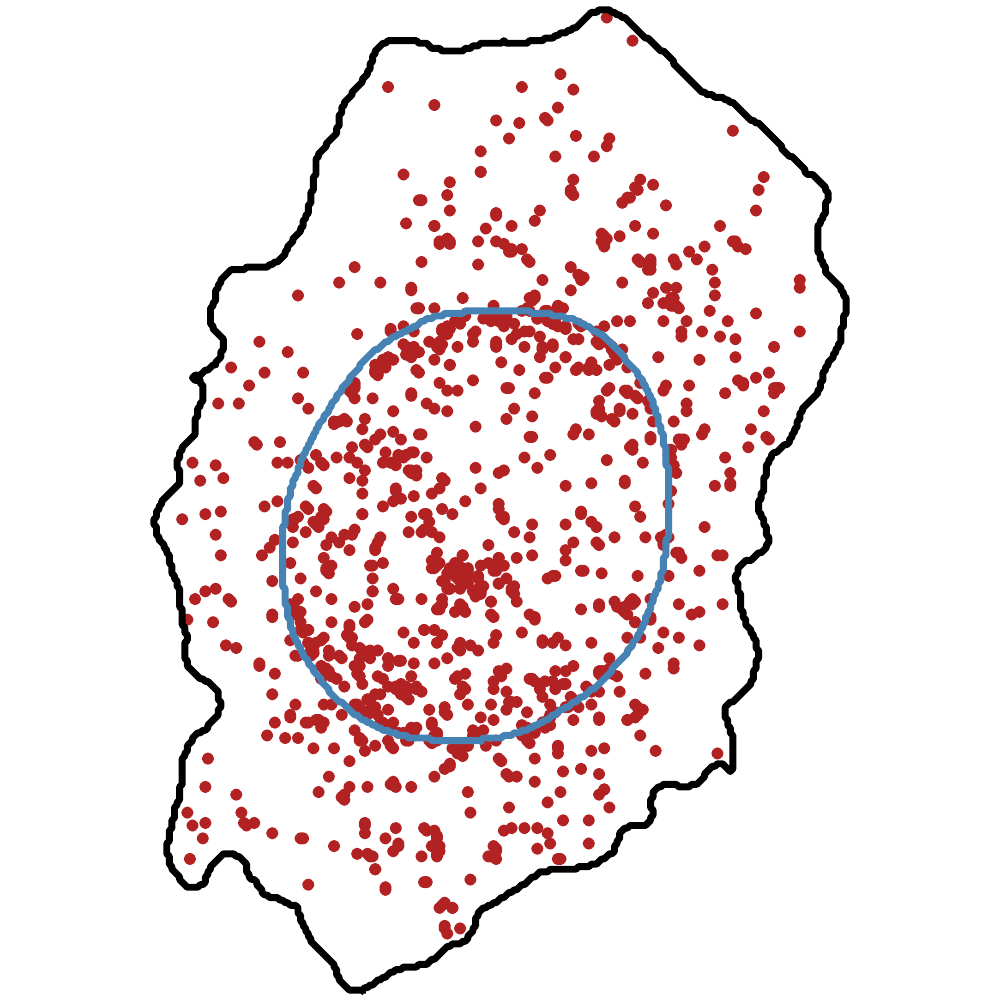}
		\subcaption{Nuclear edge}
	\endminipage\hfill
	\minipage{0.2\textwidth}
		\includegraphics[width=0.95\linewidth]{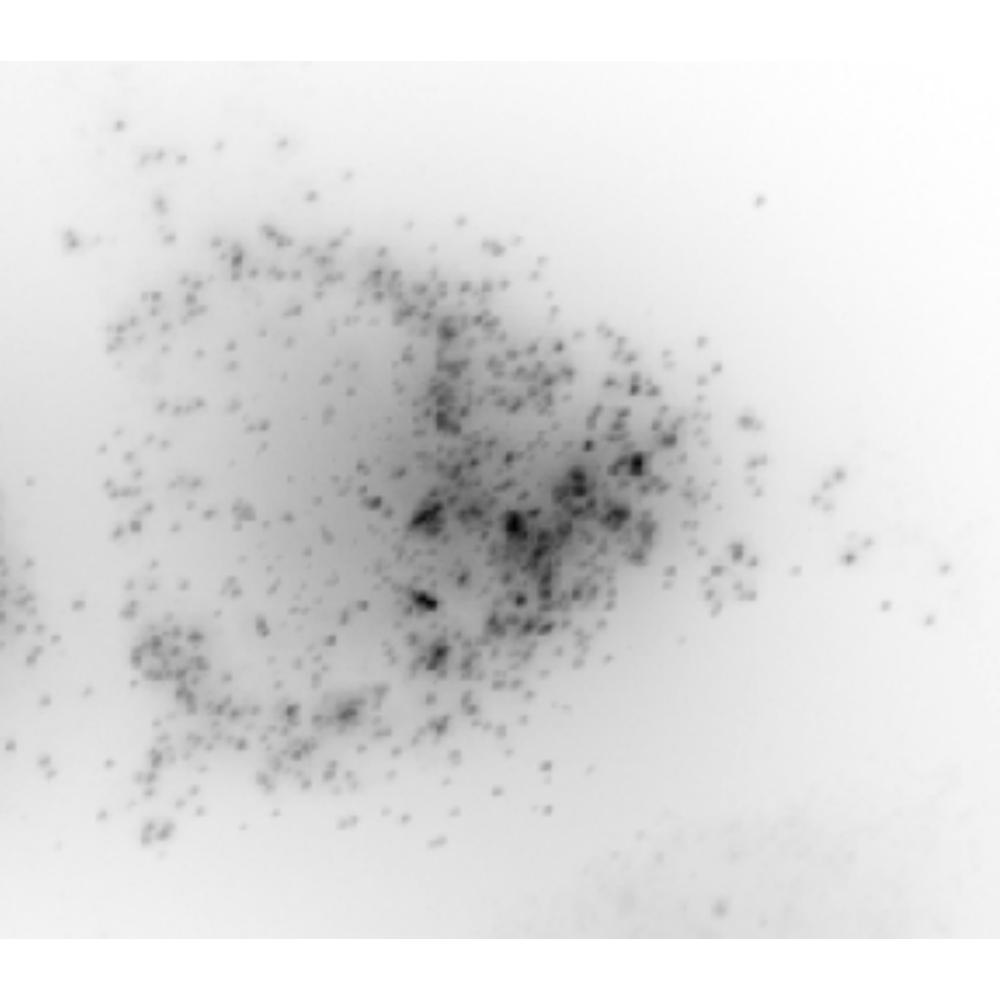}
		\vfill
		\includegraphics[width=0.95\linewidth]{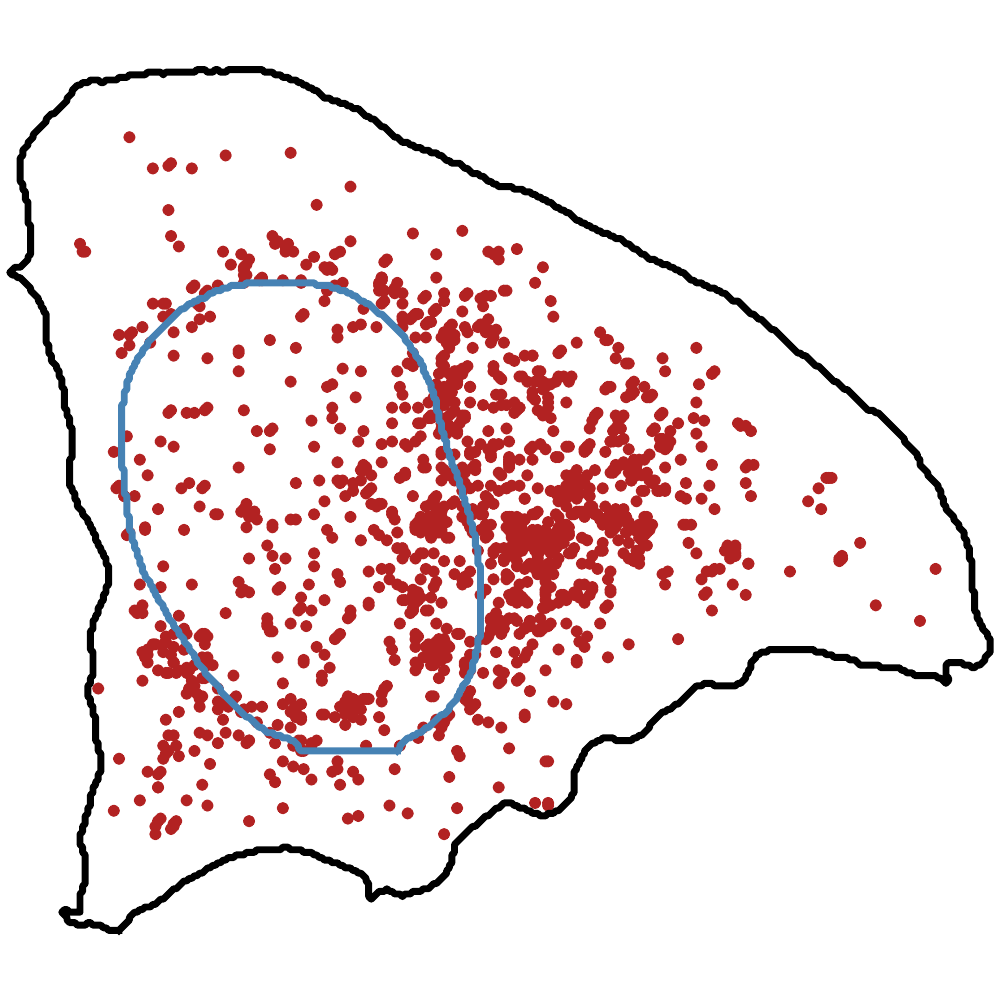}
		\subcaption{Perinuclear}
	\endminipage\hfill
	\minipage{0.2\textwidth}
		\includegraphics[width=0.95\linewidth]{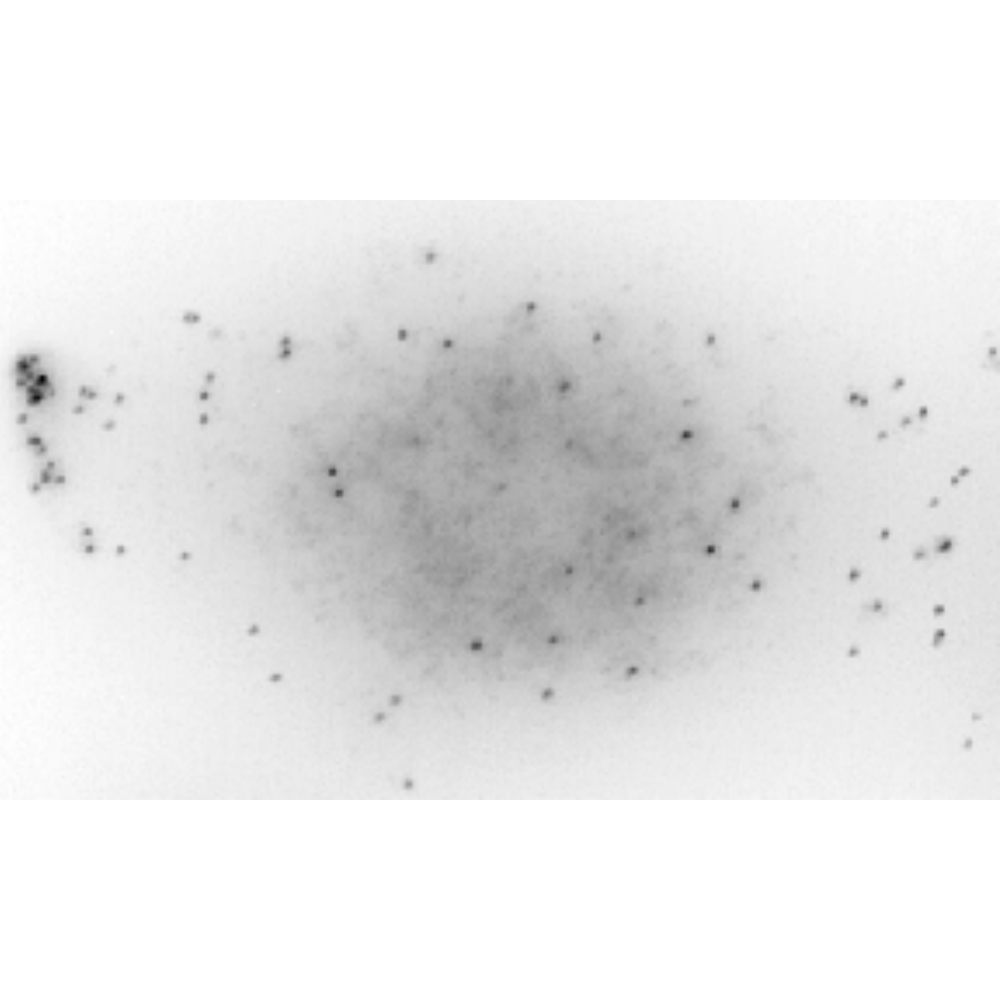}
		\vfill
		\includegraphics[width=0.95\linewidth]{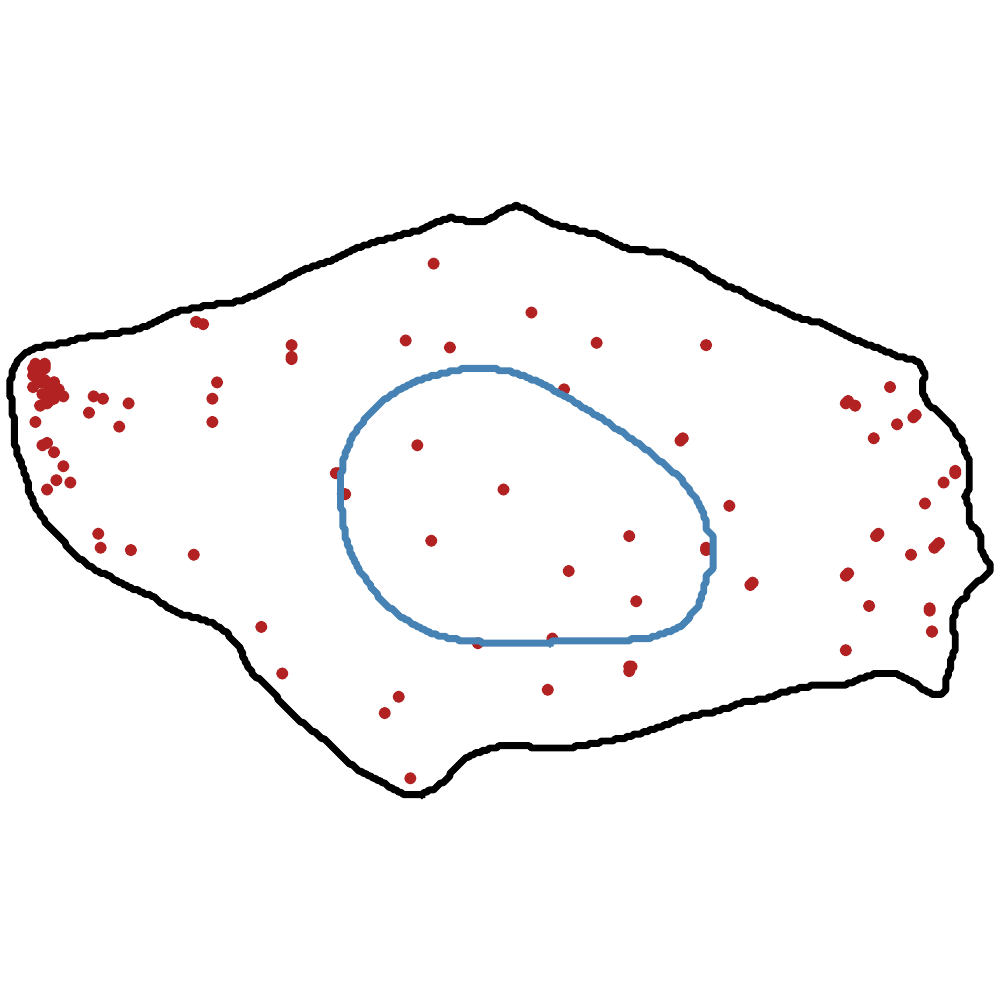}
		\subcaption{Protrusion}
	\endminipage
	\caption{RNA localization patterns from~\cite{CHOUAIB_2020}. (\textit{Top}) Typical smFISH images with different RNA localization patterns. (\textit{Bottom}) Coordinate representations with RNA spots (red), cell membrane (black) and nuclear membrane (blue). Detection and segmentation results are extracted and visualized with FISH-quant~\cite{imbert_2022}}
	\label{fig:localization_patterns}
\end{figure}

The spatial distribution of RNA can be investigated with sequence or image-based techniques.
We focus on the latter, since they provide substantially better spatial resolution and are therefore more suitable for the analysis of subcellular RNA localization.
The method of choice to visualize RNAs in intact cells is single molecule Fluorescence in Situ Hybridization (smFISH) that comes in many variants, such as a scalable and cost-efficient version ~\cite{Tsanov_2016}, which we will use in the following. In smFISH, individual RNA molecules of a given RNA species are targeted with several fluorescently labeled oligonucleotides and appear as bright diffraction-limited spots under a microscope and can thus been detected with traditional image analysis or computer vision methods. Usually, some additional fluorescent markers are used to label relevant cell structures (cytoplasm, nucleus, centrosomes, \dots), which allow one to determine the position of each individual RNA with respect to these landmarks. 
From such a multi-channel microscopy image, we can thus obtain a coordinate representation of each individual cell and its RNAs as illustrated in Figure~\ref{fig:localization_patterns}.

RNA localization results in several, distinct patterns, which can in general be defined by a local overcrowded subcellular region.
In the literature~\cite{CHOUAIB_2020}, several patterns have been described, even for a simple 
biological system such as HeLa cells: a random default pattern where RNAs localize uniformly within the cell, RNA clusters (foci), a high density of transcripts along the nuclear membrane (nuclear edge), inside the nucleus itself (intranuclear), in cell extensions (protrusion), or a polarization within the cell, like RNA localizing towards the nucleus (perinuclear).

It is still an open problem how to statistically classify and automatically detect RNA localization patterns and how to represent point cloud distributions. 
Previous approaches relied essentially on handcrafted features, quantitatively describing RNA spatial distribution within subcellular regions ~\cite{battich_image-based_2013,stoeger_computer_2015,samacoits_computational_2018}. 

Here, we intend to address the challenge to detect RNA localization patterns with a deep learning approach.
The idea is to replace the feature engineering problem by a training design problem.
Instead of manually crafting features, we propose a training procedure to learn generic encodings for RNA clouds inside cells allowing us to efficiently address the recognition of RNA localization patterns\footnote{Code and data are available in \url{https://github.com/Henley13/PointFISH}.}.

\section{Related Work}
\label{sec:related_work}

\subsubsection{Recognition of RNA Localization Pattern }

In previous studies, hand-crafted features to classify RNA localization patterns were developed~\cite{CHOUAIB_2020,stoeger_computer_2015,samacoits_computational_2018,battich_2013}.
Their design has been inspired by literature on spatial statistics~\cite{ripley2005spatial} and adapted from analysis pipelines for fluorescence microscopy images~\cite{lagache_statistical_2015,stueland_rdi_2019}.
Several packages already implement modules to perform smFISH analysis and compute these hand-crafted features~\cite{imbert_2022,mueller_2013,savulescu_dypfish_2019,mah_bento_2022}.
However, these approaches require to carefully design a set of features corresponding to the concrete biological question under study. 
For a different study a new set of features might be necessary.
Here, we aim to investigate a more general approach to build localization features.

\subsubsection{Learning Representations}

Neural network learn powerful representations that can often be used for transfer learning.
The idea is to pretrain a network and thereby to obtain a generic representation by solving a pretext task on a large annotated dataset, before addressing a more difficult or specific task with sometimes limited data. 
Often the representation optimized to solve the pretext task can also be useful for the more specific task.
Such a model can then be used as a feature extractor by computing features from one of its intermediate layers.
The computer vision community progressively replaces hand-crafted features~\cite{Lowe_1999,Bay_2006} by deep learning features to analyze images.
For instance, convolutional neural networks pretrained on large and general classification challenges~\cite{He_2016_CVPR,Szegedy_2016_CVPR,Tan_2019,Huang_2017_CVPR} are used as backbone or feature extractor for more complex tasks like face recognition, detection or segmentation.
The NLP community follows this trend as well with a heavy use of word embeddings~\cite{Mikolov_2013,Joulin_2016} or the more recent transformers models.
The same strategy has also been applied to graphs: node2vec~\cite{Grover_2016} learns ''task-independent representations'' for nodes in networks.

Such embeddings can be a continuous and numerical representation of a non-structured data like a text or a graph.
In spage2vec~\cite{Partel_2021}, the model learned a low dimensional embedding of local spatial gene expression (expressed as graphs). 
Authors then identified meaningful gene expression signatures by computing this embedding for tissue datasets.

\subsubsection{Convolutional Features}

Since we analyze imaging data, a first intuition would be to build a convolutional neural network to directly classify localization patterns from these fluorescent images.
Such approaches have a long tradition in the classification of subcellular protein localization patterns. 
Unlike RNAs, proteins are usually difficult to resolve at the single molecule level unless super-resolution microscopy was employed, which is not the case for this kind of studies. 
Protein localization patterns is therefore seen as a characteristic texture in the fluorescent image and thus the representation of subcellular protein localization often relies on texture and intensity features.
Initial studies~\cite{boland_automated_1998} computed a set of hand-crafted features from the microscopy image before training a classifier.
With the advent of deep learning, protein localization is now tackled with convolutional neural networks, but still framed as a texture classification problem. 
After crowdsourcing annotations for the Human Protein Atlas dataset~\cite{Uhlen_2015}, researchers trained a machine learning model (Loc-CAT) from hand-crafted features to predict subcellular localization patterns of proteins~\cite{sullivan_deep_2018}.
More recently, an online challenge~\cite{ouyang_analysis_2019} was organized, where the majority of top-ranking solutions were based on convolutional neural networks.
In summary, for protein localization the shift from hand-crafted features to learned representations allows for more accurate and robust pipelines.

A recent perspective paper~\cite{Savulescu_2021} suggests the increased use of deep learning models also for RNA localization analysis.
The authors emphasize the recent successes and flexibility of neural nets with different types of input, and therefore the possibility to design a multimodal pipeline. However, the fundamental difference to existing protein localization datasets, is that RNA molecules appear as distinguishable spots, and their modeling as a texture seems therefore suboptimal.

\subsubsection{Point Cloud Models}

We postulate that learning to classify RNA localization patterns directly from detected spot coordinates could be an efficient approach.
A point cloud has an unordered and irregular structure.
Projecting the point coordinates into images or voxels~\cite{Maturana_2015} transforms the problem as an easier vision challenge, but it comes along with some input degradations and dramatically increases the memory needed to process the sample. Also, relevant spatial information can be lost.
In case of RNA point clouds, it makes the recognition of 3D localization patterns harder~\cite{dubois_deep_2019}.

PointNet~\cite{Qi_2017_CVPR} is a seminal work that opened the way for innovative models to address shape classification.
It directly processes point clouds with shared MLPs and a max pooling layer, making the network invariant to input permutation.
However, the pooling step is the only way for the model to share information between close points, which ultimately limits its performance.
Yet, recent research dramatically improves point cloud modelling and especially the capture of local information.

PointNet++~\cite{Qi_2017} learns local geometric structures by recursively applying PointNet to different regions of the point cloud, in a hierarchical manner.
This way, local information can be conveyed through the network more efficiently.
DGCNN~\cite{Wang_2019} proposes a new EdgeConv layer where edge features are computed between a point and its neighbors.
Some models propose to adapt convolutions to point clouds by designing new weighting functions or kernel operations like PointCNN~\cite{Li_2018}, PointConv~\cite{Wu_2019_CVPR} or KPConv~\cite{Thomas_2019_ICCV}.
Another inspiration from the computer vision or NLP literature is the attention-based model.
To this end, PointTransformer~\cite{Zhao_2021_ICCV} proposes an attention layer to be applied to local regions within the point cloud.
Finally, PointMLP~\cite{ma2022rethinking} proposes a simple but efficient network with a pure deep hierarchical MLP architecture.

\section{Problem Statement}
\label{sec:datasets}

We want to train a model, where we can provide directly the point cloud coordinates as an input and compute a continuous vector representation.
This representation can then be used for classification of different RNA localization patterns.
Such a deep learning model might require a large volume of annotated data to reach a satisfying performance.
To generate such a large data sets, we used simulated data to train our point cloud model and then use it as a trained feature extractor.
Eventually we evaluate these learned features on a real dataset.


\begin{figure}[h]
	\centering
	\minipage{0.3\textwidth}
		\includegraphics[width=\linewidth]{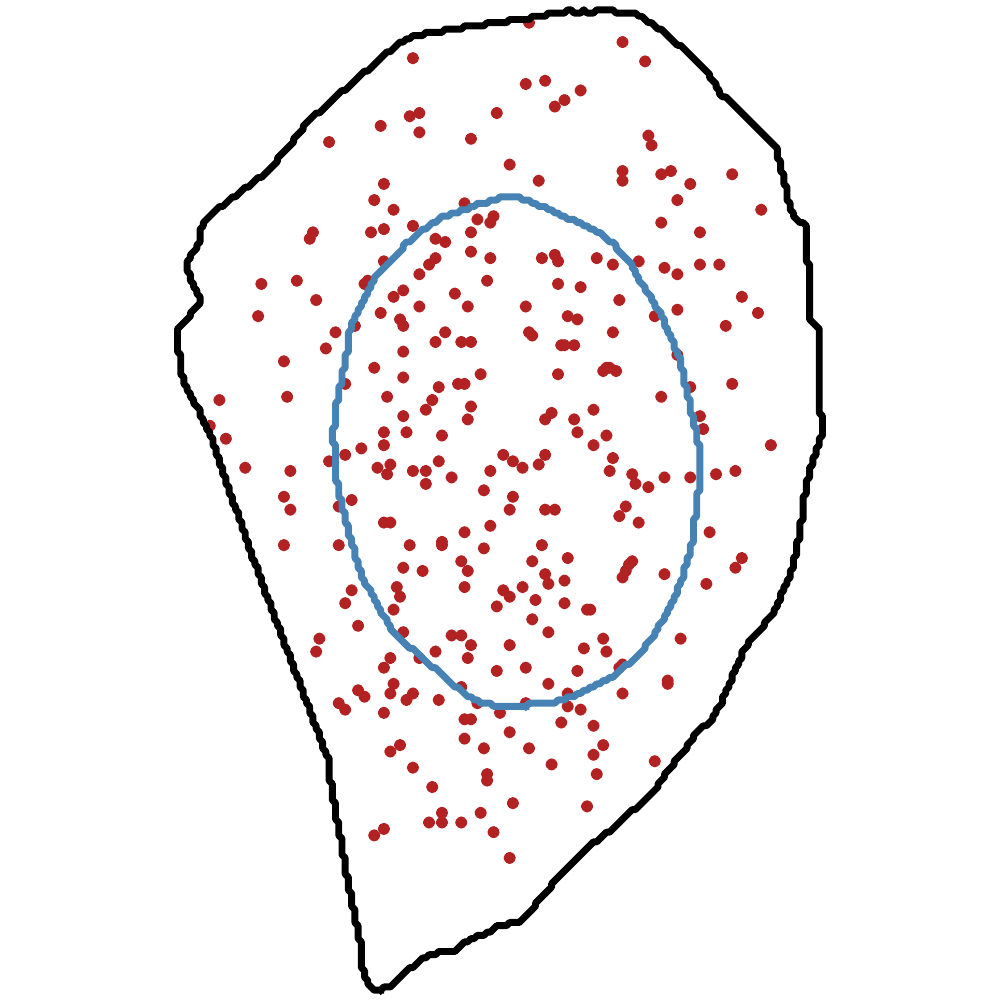}
		\subcaption{10\% perinuclear RNA}
	\endminipage\hfill
	\minipage{0.3\textwidth}
		\includegraphics[width=\linewidth]{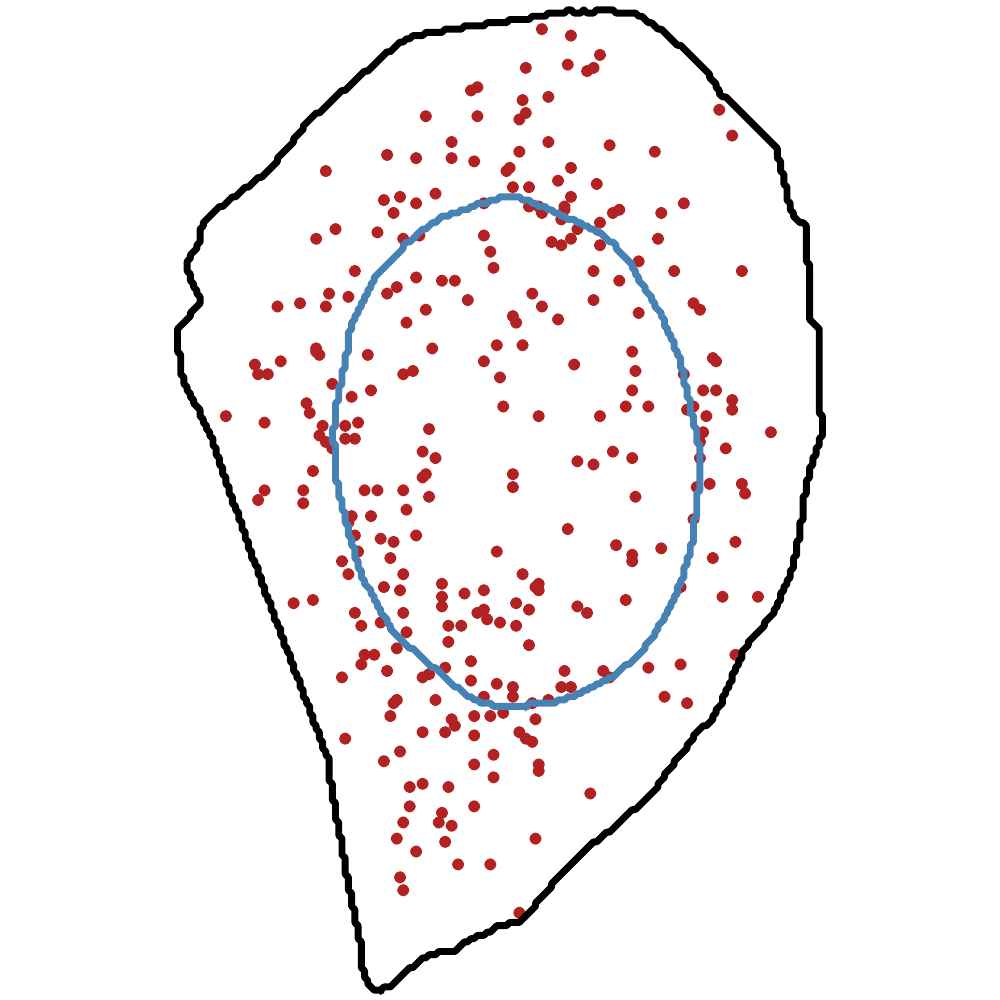}
		\subcaption{50\% perinuclear RNA}
	\endminipage\hfill
	\minipage{0.3\textwidth}
		\includegraphics[width=\linewidth]{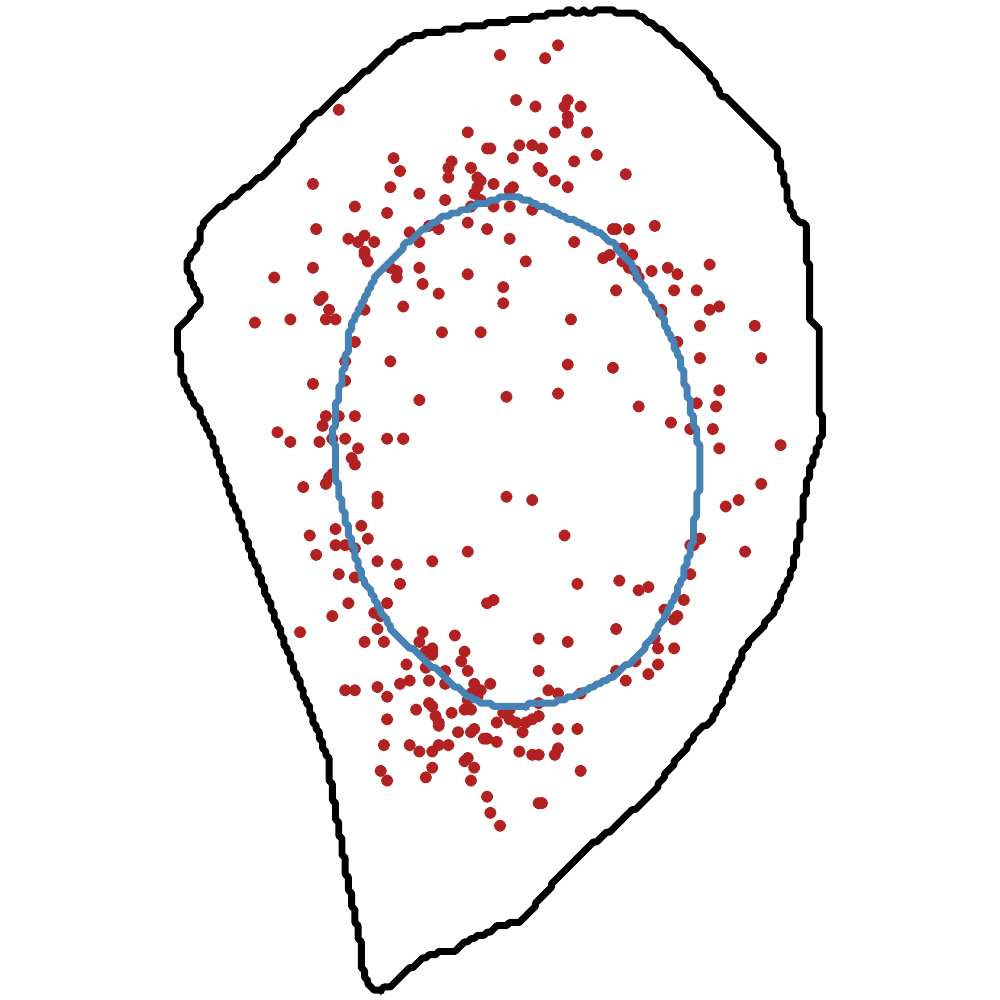}
		\subcaption{90\% perinuclear RNA}
	\endminipage
	\caption{Perinuclear pattern simulations with increasing pattern strength. Simulated with FISH-quant~\cite{imbert_2022}}
	\label{fig:perinuclear_panel}
\end{figure}

\subsubsection{Simulated Dataset}

\begin{wraptable}{R}{0.40\textwidth}
	\centering
	\caption{Annotated real dataset}
	\smallskip
	\begin{tabular}{| c | c |}
		\hline
		Pattern & \# of cells \\
		\hline
		Random & 372\\
		Foci & 198\\
		Intranuclear & 73\\
		Nuclear edge & 87\\
		Perinuclear & 64\\
		Protrusion & 83\\
		\hline
	\end{tabular}
	\label{table:real_dataset}
\end{wraptable}

Using a Python framework FISH-quant~\cite{imbert_2022}, we simulate a dataset with 8 different localization patterns: random, foci, intranuclear, extranuclear, nuclear edge, perinuclear (Figure~\ref{fig:perinuclear_panel}), cell edge and pericellular.
We choose these patterns since they represent a diverse panel of localization patterns in different cellular subregions.
We simulate for each pattern 20,000 cells with 50 to 900 RNAs per cell, resulting in a full dataset of 160,000 simulated cells.
Except for the random pattern, every simulated pattern has a proportion of RNAs with preferential localization ranging from 60\% to 100\%.
In order to test how our trained features generalize to unknown localization patterns, we deliberately omitted one of the patterns (localization in cell protrusions) from the simulation, i.e. the set of real patterns in the experimental dataset in Figure \ref{fig:localization_patterns} only partially matches the set of simulated patterns. 

We split our dataset into train, validation and test, with 60\%, 20\% and 20\% respectively.
FISH-quant simulates point clouds from a limited number of real image templates. To avoid overfitting, we make sure simulations from the same cell template can’t be assigned to different splits.
Finally, point clouds are augmented with random rotations along the up-axis, centered and divided by their absolute maximum value. This normalization step was initially performed in~\cite{Qi_2017_CVPR}.

\subsubsection{Real Dataset}

To further validate the learned feature representation on simulated images, we use a previously published experimental dataset~\cite{CHOUAIB_2020}.
These images are extracted from a smFISH study in HeLa cells targeting 27 different genes.
After data cleaning, this dataset consists of 9710 individual cells, with cropped images and coordinates extracted.
Cells have on average 346 RNAs and 90\% of them have between 39 and 1307 transcripts.
Furthermore, 810 cells have manually annotated localization patterns, as detailed in table~\ref{table:real_dataset}, providing a ground-truth for validation.
Importantly, these patterns are not mutually exclusive since cells can display several patterns at the same time, e.g. foci with a perinuclear distribution.

\section{PointFISH}
\label{sec:pointfish}

\subsection{Input Preparation}

For the simulated dataset we directly generate point clouds. 
In contrast, for any experimental dataset, the starting input is usually a multichannel fluorescent image including a smFISH channel. The first tasks are the detection of RNA molecules and the segmentation of cell and nucleus surfaces. These steps allow to retrieve for each cell a list of RNA coordinates we can format to be used as an input point cloud. For the experimental dataset we reuse the code and the extraction pipeline described in~\cite{CHOUAIB_2020}.

Besides the original RNA point cloud, we can use an optional second input vector containing additional information as input for our model.
Let $X \in \mathbb{R}^{N \times 3}$ be the original input point cloud with $N$ the number of RNAs.
We define our second input vector as $\tilde{X} \in \mathbb{R}^{N \times d}$ with $d \in \{1, 2, 3, 4, 5\}$.
It is composed of three contextual inputs: morphology input, distance input and cluster input.
First, morphological information (i.e. positional information on the plasma and nuclear membrane) is integrated by concatenating the initial RNA point cloud and  points uniformly sampled from the 2D polygons outlining the cellular and nuclear membranes.
To be consistent with a 3D point cloud input, these 2D coordinates are localized to the average height of the RNA point cloud (0 if it is centered). 
This morphological input substantially increases the size of the input point cloud $X$, because we subsample 300 points from the cell membrane and 100 points from the nuclear membrane.
To let the network discriminate between RNA, cell and nucleus points, we define two boolean vectors as contextual inputs to label the points as a cell or a nucleus point. 
A point sampled from the cell membrane will have a value (\emph{True}, \emph{False}), one from the nucleus membrane (\emph{False}, \emph{True}) and one from the original RNA point cloud (\emph{False}, \emph{False}).
We end up with $X \in \mathbb{R}^{\tilde{N} \times 3}$ (with $\tilde{N} = N + 300 + 100$) and $\tilde{X} \in \{0, 1\}^{\tilde{N} \times 2}$ as inputs.
Second, we compute the distance from cellular and nuclear membrane for each RNA in the point cloud $X$.
This adds an extra input $\tilde{X} \in \mathbb{R}^{N \times 2}$.
Third, we leverage the cluster detection algorithm from FISH-quant~\cite{imbert_2022} in order to label each RNA node as clustered or not.
It gives us a boolean $\tilde{X} \in \{0, 1\}^{N \times 1}$ to indicate if a RNA belongs to a RNA cluster of not.
Depending on whether or not we choose to add the morphological information, the distance or the clustering information, we can exploit up to 5 additional dimensions of input.

\subsection{Model Architecture}

\begin{figure}[]
    \centering
    \includegraphics[width=1\textwidth]{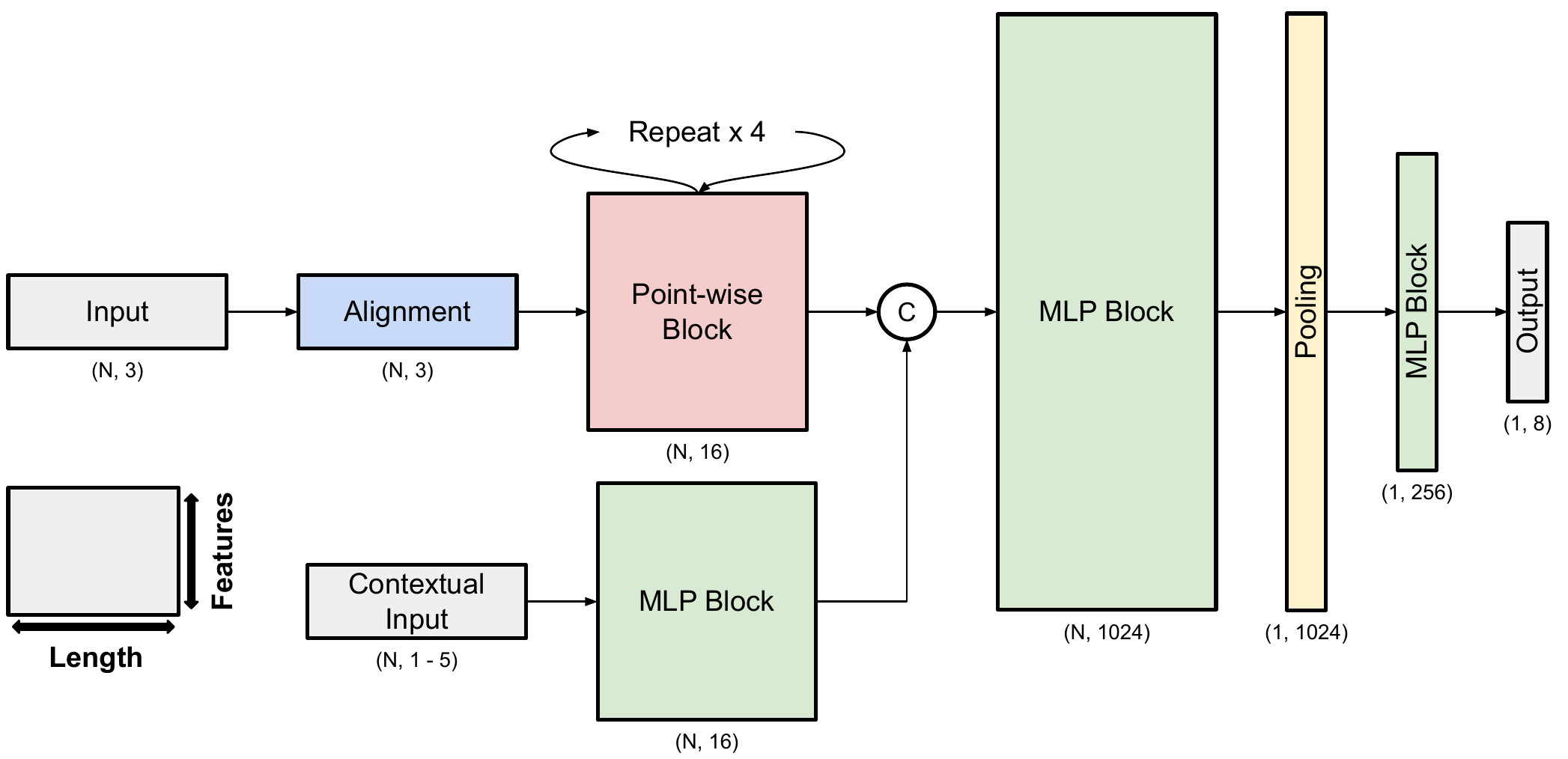}
    \caption{PointFISH architecture. Width and height of \textit{boxes} represent output length and dimension, respectively. \textit{Tuples} represent output shapes}
    \label{fig:PointFISH_architecture}
\end{figure}

We adopt the generic architecture introduced by PointNet~\cite{Qi_2017_CVPR}: successive point-wise representations with increasing depth followed by a max pooling operation to keep the network invariant by input permutation.
We incorporate state-of-the-art modules to learn efficient local structures within the point cloud.
As illustrated in Figure~\ref{fig:PointFISH_architecture}, we also adapt the network to the specificity of RNA point clouds.

\subsubsection{Point-wise Block}

Instead of shared MLPs like PointNet, we implement a multi-head attention layer based on point transformer layer~\cite{Zhao_2021_ICCV}.
First, we assign to each data point $x_i$ its 20 nearest neighbors $X(i) \subset X$, based on the euclidean distance in the feature space.
We also compute a position encoding $\delta_{ij} = \theta(x_i - x_j)$ for every pair within these neighborhoods, with $\theta$ a MLP.
Three sets of point-wise features are computed for each data point, with shared linear projections $\phi$, $\psi$ and $\alpha$.
Relative weights between data points $\gamma(\phi(x_i) - \psi(x_j))$ are computed with the subtraction relation (instead of dot product as in the seminal attention paper~\cite{NIPS2017_3f5ee243}) and a MLP $\gamma$.
These attention weights are then normalized by softmax operation $\rho$.
Eventually, data point's feature $y_i$ is computed as weighted sum of neighbors value $\alpha(x_j)$, weighted by attention.
With the position encoding added to both the attention weights and the feature value, the entire layer can be summarized such that:

\begin{align}
	{\displaystyle y_i = \sum_{x_j \in X(i)} \rho(\gamma(\phi(x_i) - \psi(x_j) + \delta_{ij})) \odot (\alpha(x_j) + \delta_{ij})}
\end{align}

For a multi-head attention layer, the process is repeated in parallel with independent layers, before a last linear projection merge multi-head outputs.
A shortcut connection and a layer normalization~\cite{ba2016layer} define the final output of our multi-head attention layer.

\subsubsection{Alignment Module}

Albeit optional (point clouds can be processed without it), this module dramatically improves performance of the network.
Some papers stress the necessity to preprocess the input point cloud by learning a projection to \emph{align} the input coordinates in the right space~\cite{Qi_2017_CVPR,Wang_2019}.
In addition, density heterogeneity across the point cloud and irregular local geometric structures might require local normalization.
To this end, we reuse the geometric affine module described in PointMLP~\cite{ma2022rethinking} which transforms local data points to a normal distribution.
With $\{x_{i, j}\}_{j=1,\dots,20} \in \mathbb{R}^{20 \times 3}$, the neighborhood's features of $x_i$, we compute:

\begin{align}
	{\displaystyle \{x_{i, j}\} = \alpha \odot \frac{\{x_{i, j}\} - x_i}{\sigma + \epsilon} + \beta}
\end{align}

\noindent
where $\alpha \in \mathbb{R}^3$ and $\beta \in \mathbb{R}^3$ are learnable parameters, $\sigma$ is the feature deviation across all local neighborhoods and $\epsilon$ is a small number for numerical stability.

\subsubsection{Contextual Inputs}

Our RNA point cloud does not include all the necessary information for a localization pattern classification.
Especially, information about the morphological properties of the cell and nucleus are lacking.
To this end, deep learning architectures allows flexible insertions.
Several contextual inputs $\tilde{X}$ can feed the network through a parallel branch, before concatenating RNA and contextual point-wise features.
Our best model exploits cluster and distance information in addition to RNA coordinates.

\section{Experiment}
\label{sec:experiment}

\subsection{Training on Simulated Patterns}

We train PointFISH on the simulated dataset.
Our implementation is based on TensorFlow~\cite{tensorflow_2015}.
We use ADAM optimizer~\cite{Diederik_2015} with a learning rate from 0.001 to 0.00001 and an exponential decay (decay rate of 0.5 every 20,000 steps).
Model is trained for a maximum of 150 epochs, with a batch size of 32, but early stopping criterion is implemented if validation loss does not decrease after 10 consecutive epochs.
Usually, the model converges after 50 epochs.
We apply a 10\% dropout for the last layer and classifications are evaluated with a categorical cross entropy loss.
Even if localization patterns are not necessarily exclusive, for the simulations we trained the model to predict only one pattern per cell.
For this reason, we did not simulate mixed patterns and assume it could help the model to learn disentangled representations.
Training takes 6 to 8 hours to converge with a Tesla P100 GPU.

A first evaluation can be performed on the simulated test dataset.
With our reference PointFISH model, we obtain a general F1-score of 95\% over the different patterns.
The configurations of this model include the use of distance and cluster contextual inputs, a geometric affine module, an attention layer as a point-wise block and a latent dimension of 256.

\subsection{Analysis of the embeddings provided by PointFISH}

From a trained PointFISH model we can remove the output layer to get a feature extractor that computes a 256-long embedding from a RNA point cloud.

\subsubsection{Exploratory analysis of experimental data embeddings}
\begin{figure}[]
    \centering
    \includegraphics[width=0.95\textwidth]{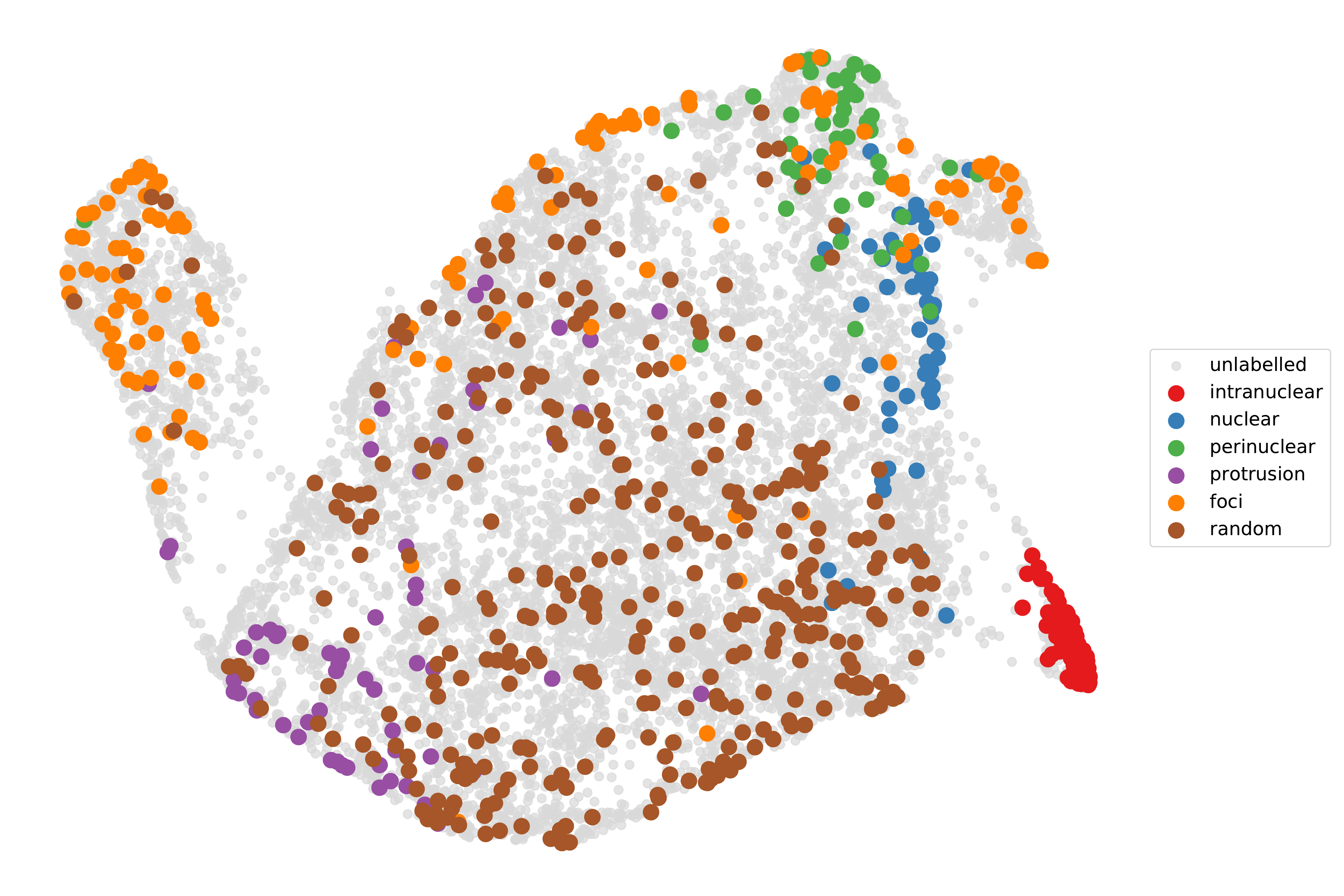}
    \caption{UMAP embedding with learned features. Each point is a cell from dataset~\cite{CHOUAIB_2020}. Manually annotated cells are colored according to their localization pattern}
    \label{fig:umap_real}
\end{figure}

We compute the embeddings for the entire cell population studied in~\cite{CHOUAIB_2020}.
All the 9170 cells can be visualized in 2D using a UMAP projection~\cite{McInnes2018}.
In Figure~\ref{fig:umap_real} each point represents a cell.
Among the 810 annotated cells, those with a unique pattern are colored according to the localization pattern observed in their RNA point cloud.
The rest of the dataset is gray.
Overall, PointFISH embedding discriminates well the different localization patterns.
Intranuclear, nuclear edge and perinuclear cells form distinct clusters, despite their spatial overlap, as well as protrusions. We recall that the protrusion patterns was not used in simulation. 
Cells with foci can be found in a separated clusters as well, but also mix with nuclear and perinuclear patterns.
This confusion is not surprising as a large number of cells in the dataset present a nuclear-related foci pattern (i.e. cells have RNAs clustered in foci, which in turn are close to the nuclear envelope). 

\subsubsection{Supervised Classification}

Because PointFISH already return meaningful embeddings, we can apply a simple classifier on top of these features to learn localization patterns.
We use the 810 manually annotated cells from the real dataset.
We compare the 15 hand-crafted features selected in~\cite{CHOUAIB_2020} with our learned embedding.
Every set of features (hand-crafted or learned) is rescaled before feeding a classifier ''by removing the mean and scaling to unit variance''\footnote{Features are rescaled with the \emph{StandardScaler} method from scikit-learn~\cite{scikit-learn}.}.
Expert features quantify RNA distributions within specific subcellular compartments and compute relevant distances to cell structures. 
Porportion features excepted, they are normalized by their expected value under a uniform RNA distribution.
Hand-crafted features include:

\begin{itemize}
	\item The number of foci and the proportion of clustered RNA.
	\item The average foci distance from nucleus and cell.
	\item The proportion or RNA inside nucleus.
	\item The average RNA distance from nucleus and cell.
	\item The number of RNAs detected in cell extensions and a peripheral dispersion index~\cite{stueland_rdi_2019}.
	\item The number of RNAs within six subcellular regions (three concentric regions around the nucleus and three others concentric regions around the cell membrane).
\end{itemize}

\begin{figure}[]
    \centering
	\includegraphics[clip, trim=0cm 0cm 0cm 1cm, width=0.9\textwidth]{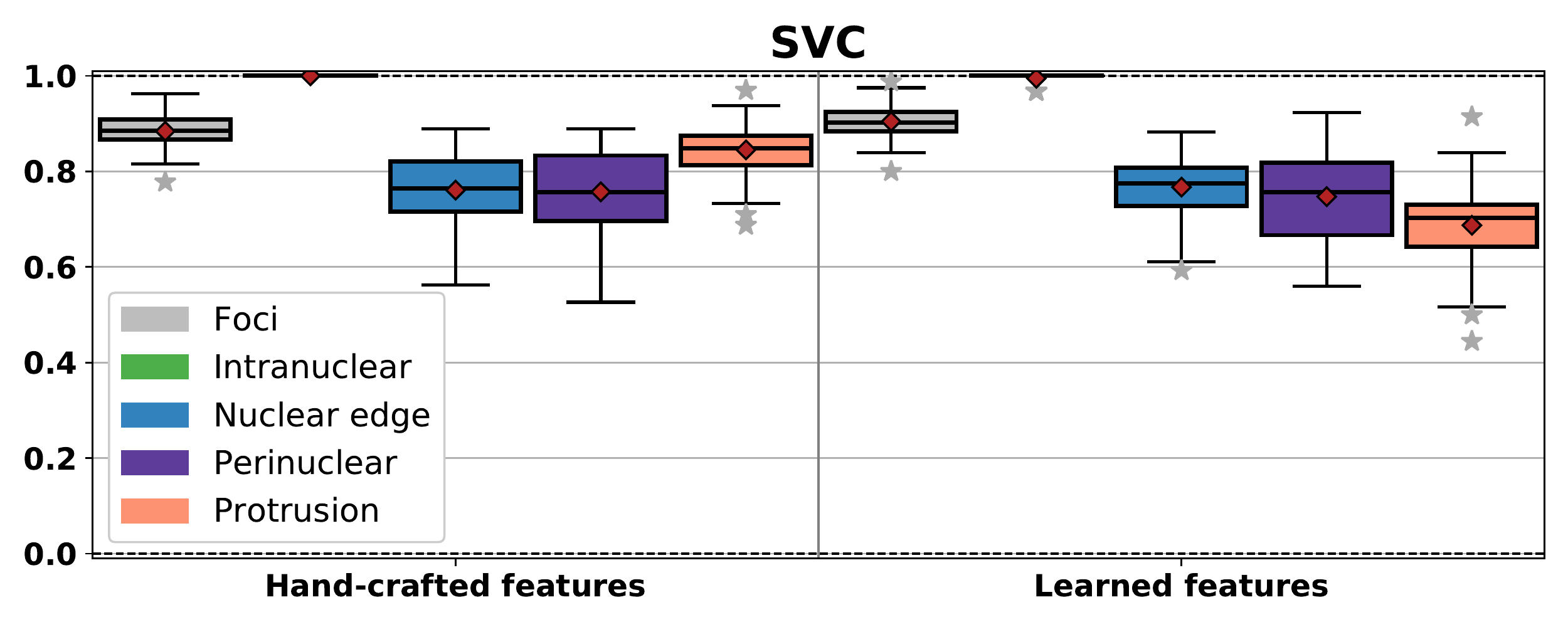}
    \caption{F1-score distribution with localization pattern classification (SVC model)}
    \label{fig:f1_SVC_real}
\end{figure}

We design 5 binary classification tasks, one per localized pattern (random pattern is omitted).
The classifier is a SVC model~\cite{chang2011libsvm}.
For evaluation purpose, we apply a nested cross-validation scheme.
First, a test dataset is sampled (20\%), then the remaining cells are used for a gridsearch to find an optimal SVC model (with another 20\% validation split).
Parameters grid includes the choice between a linear or a RBF kernel and the strength of the regularization.
The entire process is repeated 50 times, with different test split, and F1-score for each classification task is returned.
This full evaluation pipeline is implemented with scikit-learn~\cite{scikit-learn}.
F1-score's distribution over 50 splits are summarized in Figure~\ref{fig:f1_SVC_real}.
Learned features match performances of hand-crafted features selected for the tasks.
While the recognition of localization in protrusions is slightly worse, it is important to point out that we did not include simulations of this patterns in the training dataset.

\subsection{Ablation Studies}

We perform ablation studies to evaluate the impact of different components in PointFISH model.

\subsubsection{Additional Input}

\begin{wraptable}{R}{0.57\textwidth}
	\centering
	\caption{Impact of contextual inputs. F1-score is averaged over 4 trainings with different random seeds. Best model is in bold. Reference model is labelled with $\ast$}
	\smallskip
	\begin{tabular}{| c | c | c | c |}
		\hline
		Distance & Cluster & Morphology & F1-score \\
		\hline
		\ding{55} & \ding{55} & \ding{55} & 0.42 ($\pm$ 0.01)\\
		\checkmark & \ding{55} & \ding{55} & 0.74 ($\pm$ 0.02)\\
		\ding{55} & \checkmark & \ding{55} & 0.45 ($\pm$ 0.04)\\
		\checkmark & \checkmark & \ding{55} & $0.81^{\ast}$ ($\pm$ 0.01)\\
		\checkmark & \checkmark & \checkmark & \textbf{0.82} ($\pm$ 0.00)\\
		\hline
	\end{tabular}
	\label{table:extra_inputs}
\end{wraptable}

We compare the use of RNA point cloud as unique input or the inclusion of contextual information through a parallel branch.
RNA coordinates do not carry any morphological information about the cell.
In table~\ref{table:extra_inputs}, this design logically returns the lowest F1-score.
Three additional inputs are available: RNA distance from cell and nucleus (\emph{distance}), RNA clustering flag (\emph{cluster}) and the integration of cell and nucleus membrane coordinates (\emph{morphology}).
Both morphology and distance inputs can be added to provide additional information about cell morphology to the network.
However, best performances are reached when using at least distance and cluster information.
Cell and nucleus coordinates do not increase significantly the classification and dramatically increase the computation time of the model (we need to process a larger point cloud).
In particular, cluster information greatly improves the recognition of the foci pattern while morphological distances boost others localization patterns.

\subsubsection{Alignment Module and Point-wise Block}

To measure the impact of the geometric affine module~\cite{ma2022rethinking}, we compare it with the TNet module implemented in PointNet~\cite{Qi_2017_CVPR}.
We also design a variant of TNetEdge where MLP layers extracting point-wise independent features are replaced with EdgeConv layers~\cite{Wang_2019}.
Results are reported in table~\ref{table:ablation}.
An alignment block seems critical at the beginning of the network.
In addition, geometric affine module is both more efficient (F1-score of 0.81) and much lighter than TNet and TNetEdge.

Inspired by PointNet and DGCNN, we also compare the use of their respective point-wise blocks with our multi-head attention layer.
As expected, EdgeConv blocks convey a better information than PointNet by exploiting local neighborhood within point cloud (F1-score of 0.78 and 0.75 respectively).
Yet, they do not match the performance of multi-head attention layer.

Concerning these layers, we evaluate how the number of parallel heads can influence the performance of PointFISH.
By default, we use 3 parallel attention heads to let the model specialize its attentions vectors, but we also test 1, 6 and 9 parallel heads.
In table~\ref{table:ablation}, we only observe a slight benefit between the original point transformer layer~\cite{Zhao_2021_ICCV} (with one attention head) and its augmented implementations.

\subsubsection{Latent Dimensions}

The second part of PointFISH architecture is standardized: a first MLP block, a max pooling operation, a second MLP block and the output layer.
We quantify the impact of additional MLP layers within these blocks.
Our reference model returns an embedding with 256 dimensions (before the output layer).
In a MLP block, we use ReLU activation and layer normalization, but also increase or decrease the depth by a factor 2 between layers.
Before the pooling layer, the first MLP block includes 4 layers with an increasing depth (128, 256, 512 and 1024).
After the pooling layer, the second MLP block includes 2 layers with a decreasing depth (512 and 256).
Similarly, to return 128, 64 or 32 long embeddings, we implement 6 (128, 256, 512, pooling, 256 and 128), 5 (128, 256, pooling, 128 and 64) or 4 final layers (128, pooling, 64 and 32).
We observe in table~\ref{table:ablation} a reduction in performance for the lowest dimensional embedding (64 and 32).
This hyperparameter is also critical to design lighter models, with a division by 4 in terms of trainable parameters between a 256 and a 128 long embedding.

\begin{table}[]
	\centering
	\caption{Ablation studies on real dataset~\cite{CHOUAIB_2020}. F1-score is averaged over 4 trainings with different random seeds. Best models are bold. Reference model is labelled with $\ast$}
	\smallskip
	\begin{tabular}{| c | c | c | c | c | c |}
		\hline
		Alignment & Point-wise block & \# heads & \# dimensions & \# parameters & F1-score \\
		\hline\hline
		- & Attention layer & 3 & 256 & 1,372,608 & 0.73 ($\pm$ 0.00)\\
		TNet & Attention layer & 3 & 256 & 1,712,521 & 0.74 ($\pm$ 0.02)\\
		TNetEdge & Attention layer & 3 & 256 & 1,589,321 & 0.74 ($\pm$ 0.01)\\
		\hline\hline
		Affine & MLP & - & 256 & 1,374,526 & 0.75 ($\pm$ 0.01)\\
		Affine & EdgeConv & - & 256 & 1,387,006 & 0.78 ($\pm$ 0.01)\\
		\hline\hline
		Affine & Attention layer & 9 & 256 & 1,403,334 & \textbf{0.82} ($\pm$ 0.01)\\
		Affine & Attention layer & 6 & 256 & 1,387,974 & \textbf{0.82} ($\pm$ 0.01)\\
		Affine & Attention layer & 3 & 256 & 1,372,614 & $0.81^{\ast}$ ($\pm$ 0.01)\\
		Affine & Attention layer & 1 & 256 & 1,362,374 & 0.81 ($\pm$ 0.01)\\
		\hline\hline
		Affine & Attention layer & 3 & 128 & 352,966 & 0.81 ($\pm$ 0.01)\\
		Affine & Attention layer & 3 & 64 & 97,094 & 0.77 ($\pm$ 0.00)\\
		Affine & Attention layer & 3 & 32 & 32,646 & 0.75 ($\pm$ 0.01)\\
		\hline
	\end{tabular}
	\label{table:ablation}
\end{table}

\section{Discussion}
\label{sec:discussion}

We have presented a generic method of quantifying RNA localization patterns operating directly on the extracted point coordinates, without the need to design handcrafted features. 
For this, we leverage coordinates of simulated localization patterns to train a specifically designed neural network taking as input a list of points and associated features that greatly enhance generalization capabilities. 
We show that this method is on par with carefully designed, handcrafted feature sets. 

Being able to directly process list of points provides the community with a tool to integrate large datasets obtained with very different techniques on different model systems. 
While the actual image data might look strikingly different between such projects, they can all be summarized by segmentation maps of nuclei and cytoplasm, and a list of coordinates of RNA locations.
Having methods that operate directly on point clouds is therefore a strategic advantage. 

The idea of training on simulated data provides us the opportunity to query datasets with respect to new localization patterns that have not yet been observed, and for which we do not have real examples so far. 
In addition, this strategy allows us to control for potential confounders, such as cell morphology, or number of RNAs. 
Here, we provide a generic method that can leverage these simulations, without the tedious process of handcrafting new features. 
Of note, it is not necessary that the simulated patterns are optimized as to resemble real data: they rather serve as a pretext task. 
If a network is capable of distinguishing the simulated patterns, chances are high that the corresponding representation is also informative for slightly or entirely different patterns, in the same way as representations trained on ImageNet can be used for tumor detection in pathology images.
We show this by omitting the protrusion pattern from the simulation. 
We see in Figure~\ref{fig:umap_real} that the protrusion patterns live in a particular region of the feature space, without specific training. 
Moreover, we see in Figure~\ref{fig:umap_real}, that the overall separation between patterns in this exploratory way coincides to a large extent with the figure that has been proposed by the authors of the original paper~\cite{CHOUAIB_2020}. 

\section{Conclusion}
\label{sec:conclusion}

In this work, we introduce a new approach for the quantification and classification of RNA localization patterns.
On the top of existing solutions to extract RNA spots and cell morphology coordinates, we propose to directly process the resulting point clouds.
Recent advances in point cloud analysis through deep learning models allows us to build a flexible and scalable pipeline that matches results obtained with specific hand-crafted features.

Overall, with the increasing interest on subcellular RNA localization in the field of spatial transcriptomics, we expect that this approach will be of great use to the scientific community, and that it will contribute to the deciphering of some of the most fundamental processes in life.

\subsubsection{Acknowledgments}
This work was funded by the ANR (ANR-19-CE12-0007) and by the French government under management of Agence Nationale de la Recherche as part of the “Investissements d’avenir” program, reference ANR-19-P3IA-0001 (PRAIRIE 3IA Institute). Furthermore, we also acknowledge France-BioImaging infrastructure supported by the French National Research Agency (ANR-10-INBS-04).

\bibliographystyle{splncs}
\bibliography{references}

\begin{thebibliography}{10}

\bibitem{lecuyer_global_2007}
Lécuyer, E., Yoshida, H., Parthasarathy, N., Alm, C., Babak, T., Cerovina, T.,
  Hughes, T.R., Tomancak, P., Krause, H.M.:
\newblock Global analysis of {mRNA} localization reveals a prominent role in
  organizing cellular architecture and function.
\newblock Cell \textbf{131}(1) (2007)  174--187

\bibitem{buxbaum_right_2015}
Buxbaum, A.R., Haimovich, G., Singer, R.H.:
\newblock In the right place at the right time: visualizing and understanding
  {mRNA} localization.
\newblock Nature Reviews Molecular Cell Biology \textbf{16}(2) (2015)  95--109

\bibitem{Tsanov_2016}
Tsanov, N., Samacoits, A., Chouaib, R., Traboulsi, A.M., Gostan, T., Weber, C.,
  Zimmer, C., Zibara, K., Walter, T., Peter, M., Bertrand, E., Mueller, F.:
\newblock {smiFISH and FISH-quant – a flexible single RNA detection approach
  with super-resolution capability}.
\newblock Nucleic Acids Research \textbf{44}(22) (2016)  e165--e165

\bibitem{CHOUAIB_2020}
Chouaib, R., Safieddine, A., Pichon, X., Imbert, A., Kwon, O.S., Samacoits, A.,
  Traboulsi, A.M., Robert, M.C., Tsanov, N., Coleno, E., Poser, I., Zimmer, C.,
  Hyman, A., {Le Hir}, H., Zibara, K., Peter, M., Mueller, F., Walter, T.,
  Bertrand, E.:
\newblock A dual protein-mrna localization screen reveals compartmentalized
  translation and widespread co-translational rna targeting.
\newblock Developmental Cell \textbf{54}(6) (2020)  773--791.e5

\bibitem{imbert_2022}
Imbert, A., Ouyang, W., Safieddine, A., Coleno, E., Zimmer, C., Bertrand, E.,
  Walter, T., Mueller, F.:
\newblock {FISH}-quant v2: a scalable and modular tool for {smFISH} image
  analysis.
\newblock RNA \textbf{10}(6) (2022)  786--795

\bibitem{battich_image-based_2013}
Battich, N., Stoeger, T., Pelkmans, L.:
\newblock Image-based transcriptomics in thousands of single human cells at
  single-molecule resolution.
\newblock Nature Methods \textbf{10}(11) (November 2013)  1127--1133 Number: 11
  Publisher: Nature Publishing Group.

\bibitem{stoeger_computer_2015}
Stoeger, T., Battich, N., Herrmann, M.D., Yakimovich, Y., Pelkmans, L.:
\newblock Computer vision for image-based transcriptomics.
\newblock Methods \textbf{85} (September 2015)  44--53

\bibitem{samacoits_computational_2018}
Samacoits, A., Chouaib, R., Safieddine, A., Traboulsi, A.M., Ouyang, W.,
  Zimmer, C., Peter, M., Bertrand, E., Walter, T., Mueller, F.:
\newblock A computational framework to study sub-cellular {RNA} localization.
\newblock Nature Communications \textbf{9}(1) (2018)  4584

\bibitem{battich_2013}
Battich, N., Stoeger, T., Pelkmans, L.:
\newblock Image-based transcriptomics in thousands of single human cells at
  single-molecule resolution.
\newblock Nature Methods \textbf{10}(11) (2013)  1127--1133

\bibitem{ripley2005spatial}
Ripley, B.:
\newblock Spatial Statistics.
\newblock Wiley Series in Probability and Statistics. Wiley (2005)

\bibitem{lagache_statistical_2015}
Lagache, T., Sauvonnet, N., Danglot, L., Olivo-Marin, J.C.:
\newblock Statistical analysis of molecule colocalization in bioimaging.
\newblock Cytometry Part A \textbf{87}(6) (2015)  568--579

\bibitem{stueland_rdi_2019}
Stueland, M., Wang, T., Park, H.Y., Mili, S.:
\newblock {RDI} {Calculator}: {An} {Analysis} {Tool} to {Assess} {RNA}
  {Distributions} in {Cells}.
\newblock Scientific Reports \textbf{9}(1) (2019)  8267

\bibitem{mueller_2013}
Mueller, F., Senecal, A., Tantale, K., Marie-Nelly, H., Ly, N., Collin, O.,
  Basyuk, E., Bertrand, E., Darzacq, X., Zimmer, C.:
\newblock {FISH}-quant: automatic counting of transcripts in {3D} {FISH}
  images.
\newblock Nature Methods \textbf{10}(4) (2013)  277--278

\bibitem{savulescu_dypfish_2019}
Savulescu, A.F., Brackin, R., Bouilhol, E., Dartigues, B., Warrell, J.H.,
  Pimentel, M.R., Dallongeville, S., Schmoranzer, J., Olivo-Marin, J.C., Gomes,
  E.R., Nikolski, M., Mhlanga, M.M.:
\newblock {DypFISH}: {Dynamic} {Patterned} {FISH} to {Interrogate} {RNA} and
  {Protein} {Spatial} and {Temporal} {Subcellular} {Distribution} (2019)
  https://www.biorxiv.org/content/10.1101/536383v1.

\bibitem{mah_bento_2022}
Mah, C.K., Ahmed, N., Lam, D., Monell, A., Kern, C., Han, Y., Cesnik, A.J.,
  Lundberg, E., Zhu, Q., Carter, H., Yeo, G.W.:
\newblock Bento: {A} toolkit for subcellular analysis of spatial
  transcriptomics data (2022)
  https://www.biorxiv.org/content/10.1101/2022.06.10.495510v1.

\bibitem{Lowe_1999}
Lowe, D.:
\newblock Object recognition from local scale-invariant features.
\newblock In: Proceedings of the Seventh IEEE International Conference on
  Computer Vision. (1999)  1150--1157

\bibitem{Bay_2006}
Bay, H., Tuytelaars, T., Van~Gool, L.:
\newblock Surf: Speeded up robust features.
\newblock In: Computer Vision -- ECCV 2006. (2006)  404--417

\bibitem{He_2016_CVPR}
He, K., Zhang, X., Ren, S., Sun, J.:
\newblock Deep residual learning for image recognition.
\newblock In: Proceedings of the IEEE Conference on Computer Vision and Pattern
  Recognition (CVPR). (2016)

\bibitem{Szegedy_2016_CVPR}
Szegedy, C., Vanhoucke, V., Ioffe, S., Shlens, J., Wojna, Z.:
\newblock Rethinking the inception architecture for computer vision.
\newblock In: Proceedings of the IEEE Conference on Computer Vision and Pattern
  Recognition (CVPR). (2016)

\bibitem{Tan_2019}
Tan, M., Le, Q.:
\newblock {E}fficient{N}et: Rethinking model scaling for convolutional neural
  networks.
\newblock In: Proceedings of the 36th International Conference on Machine
  Learning. (2019)  6105--6114

\bibitem{Huang_2017_CVPR}
Huang, G., Liu, Z., van~der Maaten, L., Weinberger, K.Q.:
\newblock Densely connected convolutional networks.
\newblock In: Proceedings of the IEEE Conference on Computer Vision and Pattern
  Recognition (CVPR). (2017)

\bibitem{Mikolov_2013}
Mikolov, T., Chen, K., Corrado, G., Dean, J.:
\newblock Efficient estimation of word representations in vector space (2013)
  https://arxiv.org/abs/1301.3781.

\bibitem{Joulin_2016}
Joulin, A., Grave, E., Bojanowski, P., Mikolov, T.:
\newblock Bag of tricks for efficient text classification (2016)
  https://arxiv.org/abs/1607.01759.

\bibitem{Grover_2016}
Grover, A., Leskovec, J.:
\newblock Node2vec: Scalable feature learning for networks.
\newblock In: Proceedings of the 22nd ACM SIGKDD International Conference on
  Knowledge Discovery and Data Mining. (2016)  855–864

\bibitem{Partel_2021}
Partel, G., Wählby, C.:
\newblock Spage2vec: Unsupervised representation of localized spatial gene
  expression signatures.
\newblock The FEBS Journal \textbf{288}(6) (2021)  1859--1870

\bibitem{boland_automated_1998}
Boland, M.V., Markey, M.K., Murphy, R.F.:
\newblock Automated recognition of patterns characteristic of subcellular
  structures in fluorescence microscopy images.
\newblock Cytometry \textbf{33}(3) (1998)  366--375

\bibitem{Uhlen_2015}
Uhlén, M., Fagerberg, L., Hallström, B.M., Lindskog, C., Oksvold, P.,
  Mardinoglu, A., Åsa Sivertsson, Kampf, C., Sjöstedt, E., Asplund, A.,
  Olsson, I., Edlund, K., Lundberg, E., Navani, S., Szigyarto, C.A.K., Odeberg,
  J., Djureinovic, D., Takanen, J.O., Hober, S., Alm, T., Edqvist, P.H.,
  Berling, H., Tegel, H., Mulder, J., Rockberg, J., Nilsson, P., Schwenk, J.M.,
  Hamsten, M., von Feilitzen, K., Forsberg, M., Persson, L., Johansson, F.,
  Zwahlen, M., von Heijne, G., Nielsen, J., Pontén, F.:
\newblock Tissue-based map of the human proteome.
\newblock Science \textbf{347}(6220) (2015)  1260419

\bibitem{sullivan_deep_2018}
Sullivan, D.P., Winsnes, C.F., Åkesson, L., Hjelmare, M., Wiking, M.,
  Schutten, R., Campbell, L., Leifsson, H., Rhodes, S., Nordgren, A., Smith,
  K., Revaz, B., Finnbogason, B., Szantner, A., Lundberg, E.:
\newblock Deep learning is combined with massive-scale citizen science to
  improve large-scale image classification.
\newblock Nature Biotechnology \textbf{36}(9) (2018)  820--828

\bibitem{ouyang_analysis_2019}
Ouyang, W., Winsnes, C.F., Hjelmare, M., Cesnik, A.J., Åkesson, L., Xu, H.,
  Sullivan, D.P., Dai, S., Lan, J., Jinmo, P., Galib, S.M., Henkel, C., Hwang,
  K., Poplavskiy, D., Tunguz, B., Wolfinger, R.D., Gu, Y., Li, C., Xie, J.,
  Buslov, D., Fironov, S., Kiselev, A., Panchenko, D., Cao, X., Wei, R., Wu,
  Y., Zhu, X., Tseng, K.L., Gao, Z., Ju, C., Yi, X., Zheng, H., Kappel, C.,
  Lundberg, E.:
\newblock Analysis of the {Human} {Protein} {Atlas} {Image} {Classification}
  competition.
\newblock Nature Methods \textbf{16}(12) (2019)  1254--1261

\bibitem{Savulescu_2021}
Savulescu, A.F., Bouilhol, E., Beaume, N., Nikolski, M.:
\newblock Prediction of rna subcellular localization: Learning from
  heterogeneous data sources.
\newblock iScience \textbf{24}(11) (2021)  103298

\bibitem{Maturana_2015}
Maturana, D., Scherer, S.:
\newblock Voxnet: A 3d convolutional neural network for real-time object
  recognition.
\newblock In: 2015 IEEE/RSJ International Conference on Intelligent Robots and
  Systems (IROS). (2015)  922--928

\bibitem{dubois_deep_2019}
Dubois, R., Imbert, A., Samacoïts, A., Peter, M., Bertrand, E., Müller, F.,
  Walter, T.:
\newblock A {Deep} {Learning} {Approach} {To} {Identify} {mRNA} {Localization}
  {Patterns}.
\newblock In: 2019 {IEEE} 16th {International} {Symposium} on {Biomedical}
  {Imaging} ({ISBI}). (2019)  1386--1390

\bibitem{Qi_2017_CVPR}
Qi, C.R., Su, H., Mo, K., Guibas, L.J.:
\newblock Pointnet: Deep learning on point sets for 3d classification and
  segmentation.
\newblock In: Proceedings of the IEEE Conference on Computer Vision and Pattern
  Recognition (CVPR). (2017)

\bibitem{Qi_2017}
Qi, C.R., Yi, L., Su, H., Guibas, L.J.:
\newblock Pointnet++: Deep hierarchical feature learning on point sets in a
  metric space.
\newblock In: Advances in Neural Information Processing Systems. Volume~30.
  (2017)

\bibitem{Wang_2019}
Wang, Y., Sun, Y., Liu, Z., Sarma, S.E., Bronstein, M.M., Solomon, J.M.:
\newblock Dynamic graph cnn for learning on point clouds.
\newblock ACM Trans. Graph. \textbf{38}(5) (2019)

\bibitem{Li_2018}
Li, Y., Bu, R., Sun, M., Wu, W., Di, X., Chen, B.:
\newblock Pointcnn: Convolution on x-transformed points.
\newblock In: Advances in Neural Information Processing Systems. Volume~31.
  (2018)

\bibitem{Wu_2019_CVPR}
Wu, W., Qi, Z., Fuxin, L.:
\newblock Pointconv: Deep convolutional networks on 3d point clouds.
\newblock In: Proceedings of the IEEE/CVF Conference on Computer Vision and
  Pattern Recognition (CVPR). (2019)

\bibitem{Thomas_2019_ICCV}
Thomas, H., Qi, C.R., Deschaud, J.E., Marcotegui, B., Goulette, F., Guibas,
  L.J.:
\newblock Kpconv: Flexible and deformable convolution for point clouds.
\newblock In: Proceedings of the IEEE/CVF International Conference on Computer
  Vision (ICCV). (2019)

\bibitem{Zhao_2021_ICCV}
Zhao, H., Jiang, L., Jia, J., Torr, P.H., Koltun, V.:
\newblock Point transformer.
\newblock In: Proceedings of the IEEE/CVF International Conference on Computer
  Vision (ICCV). (2021)  16259--16268

\bibitem{ma2022rethinking}
Ma, X., Qin, C., You, H., Ran, H., Fu, Y.:
\newblock Rethinking network design and local geometry in point cloud: A simple
  residual {MLP} framework.
\newblock In: International Conference on Learning Representations. (2022)

\bibitem{NIPS2017_3f5ee243}
Vaswani, A., Shazeer, N., Parmar, N., Uszkoreit, J., Jones, L., Gomez, A.N.,
  Kaiser, L.u., Polosukhin, I.:
\newblock Attention is all you need.
\newblock In: Advances in Neural Information Processing Systems. Volume~30.
  (2017)

\bibitem{ba2016layer}
Ba, J.L., Kiros, J.R., Hinton, G.E.:
\newblock Layer normalization (2016) http://arxiv.org/abs/1607.06450.

\bibitem{tensorflow_2015}
Abadi, M., Agarwal, A., Barham, P., Brevdo, E., Chen, Z., Citro, C., Corrado,
  G.S., Davis, A., Dean, J., Devin, M., Ghemawat, S., Goodfellow, I., Harp, A.,
  Irving, G., Isard, M., Jia, Y., Jozefowicz, R., Kaiser, L., Kudlur, M.,
  Levenberg, J., Man\'{e}, D., Monga, R., Moore, S., Murray, D., Olah, C.,
  Schuster, M., Shlens, J., Steiner, B., Sutskever, I., Talwar, K., Tucker, P.,
  Vanhoucke, V., Vasudevan, V., Vi\'{e}gas, F., Vinyals, O., Warden, P.,
  Wattenberg, M., Wicke, M., Yu, Y., Zheng, X.:
\newblock {TensorFlow}: Large-scale machine learning on heterogeneous systems
  (2015) https://www.tensorflow.org/.

\bibitem{Diederik_2015}
Kingma, D.P., Ba, J.:
\newblock Adam: A method for stochastic optimization.
\newblock CoRR (2015)

\bibitem{McInnes2018}
McInnes, L., Healy, J., Saul, N., Großberger, L.:
\newblock Umap: Uniform manifold approximation and projection.
\newblock Journal of Open Source Software \textbf{3}(29) (2018)  861

\bibitem{scikit-learn}
Pedregosa, F., Varoquaux, G., Gramfort, A., Michel, V., Thirion, B., Grisel,
  O., Blondel, M., Prettenhofer, P., Weiss, R., Dubourg, V., Vanderplas, J.,
  Passos, A., Cournapeau, D., Brucher, M., Perrot, M., Duchesnay, E.:
\newblock Scikit-learn: Machine learning in {P}ython.
\newblock Journal of Machine Learning Research \textbf{12} (2011)  2825--2830

\bibitem{chang2011libsvm}
Chang, C.C., Lin, C.J.:
\newblock Libsvm: A library for support vector machines.
\newblock ACM transactions on intelligent systems and technology (TIST)
  \textbf{2}(3) (2011)  1--27

\end{thebibliography}

\end{document}